\documentclass[letterpaper,times]{IONconf}
\usepackage{amsmath}
\usepackage{amssymb}
\usepackage{bm}

\usepackage{graphicx}
\usepackage[justification=centering]{caption}
\usepackage{subcaption}
\usepackage{parskip}

\usepackage{array}
\usepackage{multirow}
\usepackage{booktabs}
\usepackage{bbding} 
\usepackage{float}
\usepackage{multicol}


\usepackage[%
backend=biber,
natbib=true,
style=apa,
]{biblatex}

\usepackage{cleveref}

\title{GNSS Measurement-Based Context Recognition for Vehicle Navigation\\ using Gated Recurrent Unit}

\author{
    Sheng Liu, \textit{School of Mathematics and  Computational Science, Xiangtan University;  Xiangtan, 411105, China}
    \vspace{1mm} \\%
    Zhiqiang Yao, \textit{School of Automation and Electronic Information, Xiangtan University, Xiangtan, 411105, China}%
    \vspace{1mm} \\%
    Xuemeng Cao, \textit{School of Automation and Electronic Information, Xiangtan University, Xiangtan, 411105, China}%
    \vspace{1mm} \\%
    Xiaowen Cai, \textit{School of Automation and Electronic Information, Xiangtan University, Xiangtan, 411105, China}%
    }

\begin{document}

\maketitle

\section*{biography}


\biography{Sheng Liu}{received his B.S. degree in Electronic Information Science and Technology from Xiangtan University in June 2018. He is currently a Ph.D. candidate at the School of Mathematics and  Computational Science in Xiangtan University. His research interests include optimization algorithms in context-adaptive navigation and scenario recognition for vehicle navigation.}

\biography{Zhiqiang Yao}{received his master's and doctorate degrees in engineering from South China University of Technology in 2004 and 2010, respectively. He is currently the dean (professor) of the School of Automation and Electronic Information at Xiangtan University. He serves as the director of the Key Laboratory of Intelligent Reliable Navigation and Positioning of Hunan Province and holds the position of director at the China Satellite Navigation and Positioning Association. His research interests include multi-sensor integrated navigation and context-adaptive navigation.}

\biography{Xuemeng Cao}{ received his B.S. degree in Electronic Information Engineering from Xiangtan University in 2021. He is currently engaged in the pursuit of a master's degree in the same field. His research focuses on scenario recognition for navigation and multirate multisensor data fusion.}

\biography{Xiaowen Cai}{received the Ph.D. degree in optical engineering with Beihang University, Beijing, China, in 2020. Her research interests are related to inertial navigation and integrated
	navigation.}

\section*{Abstract}
Recent years, people have put forward higher and higher requirements for context-adaptive navigation (CAN).
CAN system realizes seamless navigation in complex environments by recognizing the ambient surroundings of vehicles, and it is crucial to develop a fast, reliable, and robust navigational context recognition (NCR) method to enable CAN systems to operate effectively. Environmental context recognition based on Global Navigation Satellite System (GNSS)  measurements has attracted widespread attention due to its low cost because it does not require additional infrastructure. The performance and application value of NCR methods depend on three main factors: context categorization, feature extraction, and classification models.  
In this paper, a fine-grained context categorization framework comprising seven environment categories (open sky, tree-lined avenue, semi-outdoor, urban canyon, viaduct-down, shallow indoor, and deep indoor) is proposed, which currently represents the most elaborate context categorization framework known in this research domain.
To improve discrimination between categories, a new feature called the C/N0-weighted azimuth distribution factor, is designed. Then, to ensure real-time performance, a lightweight gated recurrent unit (GRU) network is adopted for its excellent sequence data processing capabilities.
A dataset containing 59,996 samples is created and made publicly available to researchers in the NCR community on Github.
Extensive experiments have been conducted on the dataset, and the results show that the proposed method achieves an overall recognition accuracy of 99.41\% for isolated scenarios and 94.95\% for transition scenarios, with an average transition delay of 2.14 seconds.

\section{Introduction}
With the rapid development of autonomous driving, integrated navigation and seamless positioning \citep{Zhu2017,DoNascimento2021,Wang2022}, people have put forward higher and higher requirements for context-adaptive navigation (CAN). Vehicle navigation and positioning are highly dependent on the surrounding environments, and no single technology can adapt to all operating environments. For instance, GNSS-based navigation performs well in open environments with few or no signal obstructions and interference; in semi-open environments, especially in urban canyons, shadow matching \citep{Groves2011} and 3DMA \citep{adjrad2016intelligent} can be used to assist GNSS localization; in indoor and severe GNSS-denied environments, navigation methods based on cellular signal, WiFi, Bluetooth, UWB and INS are mainly adopted \citep{kassas2017hear,Gao2018}. CAN systems achieve seamless navigation in complex environments by recognizing the ambient surroundings of the vehicle. Therefore, it is crucial to develop a fast, reliable, and robust navigational context recognition (NCR) method to enable CAN systems' effective operation.

The effectiveness and practicality of NCR methods mainly rely on three key factors: context categorization, feature extraction, and classification model. 

First, context categorization proposed for general or other purposes may not be suitable for CAN. For specific navigation applications, such as autonomous driving, a dedicated context categorization framework should be proposed based on the navigation requirements and the characteristics of different contexts. To the best of our knowledge, the existing literature \citep{zhu2020indoor,xia2020recurrent,Dai2022} mainly divides the environments under consideration into four categories or less, namely deep indoor, shallow indoor, semi-outdoor, and open-sky. Such context categorization methods are too coarse for vehicle navigation to be used in CAN. 
Second, selecting the appropriate signal or sensor types and extracting the proper features are the cornerstone of NCR. Among the various types of sensors or signals for NCR, the recognition scheme based on GNSS signals is widely used due to its universality, high availability, and low cost \citep{Feriol2020}. In terms of identification features, researchers have proposed a large number of useful features to distinguish various types of context, such as the mean and variance of GNSS signal's carrier-to-noise ratio ($ C/N_0 $), satellite blocking coefficient, and fluctuation coefficient \citep{wang2019urban}. However, when more elaborate classifications are required, new features need to be devised. 
Last, determining the appropriate classification model is crucial to achieve fast, accurate, and robust context recognition performance. Existing classification models mainly include fuzzy inference \citep{zadeh1996fuzzy}, support vector machine (SVM) \citep{suthaharan2016support}, and long-short term memory (LSTM) \citep{sherstinsky2020fundamentals}. With the development of big data technology, large-scale parallel computing, and the popularity of graphics processing unit (GPU) devices, deep learning has emerged as a promising field, with algorithms such as Convolutional Neural Networks (CNN, \cite{Dai2022}), Transformers \citep{vaswani2017attention} and Gated Recurrent Unit (GRU, \cite{chung2014empirical}).

For the GNSS measurement-based methods, a classical approach is called SatProbe \citep{8057095}, which determines the indoor/outdoor status using only the number of visible GPS satellites. Although this method turn out to be efficient, its binary classification needs improvement for applicability in broader contexts. \citet{xia2020recurrent} used an LSTM network, to divide environments into four categories (deep indoors, shallow indoors, semi-outdoors, and open outdoors) based on smartphone GNSS measurements. They achieved an overall accuracy of 98.65\% and a maximum scenario transition recognition delay of 3s. However, the constructed features suffered from redundancy, leading to unnecessary computations. Moreover, most smartphones' GNSS modules have a sampling rate of only 1Hz, which hampers the responsiveness to scenario transition. More recently, \citet{Dai2022} proposed a grid-based recognition approach that utilizes GNSS measurements such as pseudorange, Doppler shift, and C/N0. They represented the GNSS measurements with Voronoi diagrams and fed them into CNN networks, and achieved an accuracy of 99.92\%. However, the categorization is not specifically designed for vehicle navigation. In addition, classifying images generally incurs higher computational overhead compared to numerical samples.

To summarize, NCR faces three main challenges, namely categorization framework, feature extraction, and classification models. In response to these challenges, we dedicate to propose an elaborate categorization framework, and implement recognition by using a lightweight network with appropriate features.
The main contributions of this paper are:
\begin{itemize}		
	\item[$\bullet$] A novel fine-grained context categorization framework was proposed based on the characteristics of different environments and their corresponding integrated navigation methods, which currently represents the most elaborate context categorization framework known in this research field.
	\item[$\bullet$] To improve discrimination between categories, a new feature called the C/N0-weighted azimuth distribution factor was designed.
	\item[$\bullet$] To ensure real-time performance, a lightweight GRU network was adopted for its excellent sequence data processing capabilities.
	\item[$\bullet$] A corresponding dataset containing 59,996 samples was created, which will serve as a valuable resource for the NCR research community.
\end{itemize}

\section{methodology}

\vspace{-.2cm}
\subsection{Context Categorization Framework}
Targeting the navigation of autonomous vehicles in urban areas, the environments that they traverse are diverse. The selection of environmental category elements should be based on the available sensor types and the quality of wireless signal reception. The most used four-category framework (deep indoors, shallow indoors, semi-outdoors, and open outdoors) provide a basic skeleton, but it's not dedicated designed for vehicle navigation. We expand it by  adding three other distinct categories and  propose a  fine-grained context categorization framework  based on the characteristics of different environments and their corresponding integrated navigation methods.  

Table \ref{tab:my-table}  shows the definitions, signal characteristics, and corresponding integrated navigation methods for each environmental types in the proposed seven-category framework.
It includes seven categories: open sky, tree-lined avenues, semi-outdoor, urban canyons, viaduct-down, shallow indoor, and deep indoor, which currently represents the most elaborate context categorization framework known in this research domain.
For example, the simi-outdoor environment refers to a location where there is a building exists on one side and the elevation of its top edge is greater than $45^\circ$. Signals coming from elevations lower than this angle on that side are blocked. The navigation solution for the semi-outdoor environment can be achieved by the integration of GNSS, WiFi, cellular signals and/or INS.

The seven-category framework significantly improves the environmental coverage, but it also increases the difficulty of inter-class discrimination. For example, in the four-category framework, there is no category for viaduct-down, and according to the signal characteristics, it should be classified as a shallow indoor environment. However, how to distinguish between viaduct-down and shallow indoor in the seven-category framework? New features need to be designed to solve the inter-class confusion problem brought by the extended framework.

\begin{table}[]
	\vspace*{-.3cm}
	\centering
	\caption{Context categorization framework based on the characteristics of different environments and their corresponding integrated navigation methods.}
	\label{tab:my-table}
	\resizebox{\textwidth}{!}{%
		\begin{tabular}{@{}cccc@{}}
			\toprule
			categories &
			descriptions &
			GNSS signal characteristics &
			integrated navigation methods \\ \midrule
			open sky &
			\begin{tabular}[c]{@{}c@{}}position with   an elevation angle of\\      less than $ 15^\circ $  from the top edge   of\\      most surrounding buildings\end{tabular} &
			\begin{tabular}[c]{@{}c@{}}all   GNSS signals in the sky are well\\      received with strong $ C/N_0 $\end{tabular} &
			GNSS \\
			tree-lined   avenue &
			\begin{tabular}[c]{@{}c@{}}leaves on both   sides of the road\\      almost cover the sky above the road\end{tabular} &
			\begin{tabular}[c]{@{}c@{}}signal   strength experiences some\\      degree of attenuation\end{tabular} &
			GNSS, LTE/5G, INS \\
			semi-outdoor &
			\begin{tabular}[c]{@{}c@{}}building   exists on one side and the\\      elevation of its top edge is greater\\      than $ 45^\circ $ \end{tabular} &
			\begin{tabular}[c]{@{}c@{}}signals at low and medium elevation\\      angles on one side are blocked\end{tabular} &
			GNSS, WiFi, LTE/5G, INS \\
			urban   canyon &
			\begin{tabular}[c]{@{}c@{}}a large number of super high-rise\\      buildings on both sides of the road\end{tabular} &
			\begin{tabular}[c]{@{}c@{}}signals come from high elevation   on\\      the top\end{tabular} &
			\begin{tabular}[c]{@{}c@{}}GNSS Shadow matching, \\      3DMA, INS\end{tabular} \\
			viaduct-down &
			position located under a viaduct &
			\begin{tabular}[c]{@{}c@{}}signals come from low elevation   on\\      both sides\end{tabular} &
			GNSS, UWB, INS \\
			shallow   indoor &
			\begin{tabular}[c]{@{}c@{}}position located indoors and   near an\\      outside window\end{tabular} &
			\begin{tabular}[c]{@{}c@{}}weak GNSS signals coming from   the\\      side of the window\end{tabular} &
			GNSS, WiFi, INS \\
			deep   indoor &
			\begin{tabular}[c]{@{}c@{}}position located indoors and away\\ from an outside window\end{tabular} &
			almost no GNSS signal available &
			WiFi, Bluetooth, UWB, INS \\ \bottomrule
		\end{tabular}%
	}
\end{table}
\subsection{Feature Design/Selection}
\begin{figure}[b]
	\centering
	\includegraphics[width=.9\linewidth]{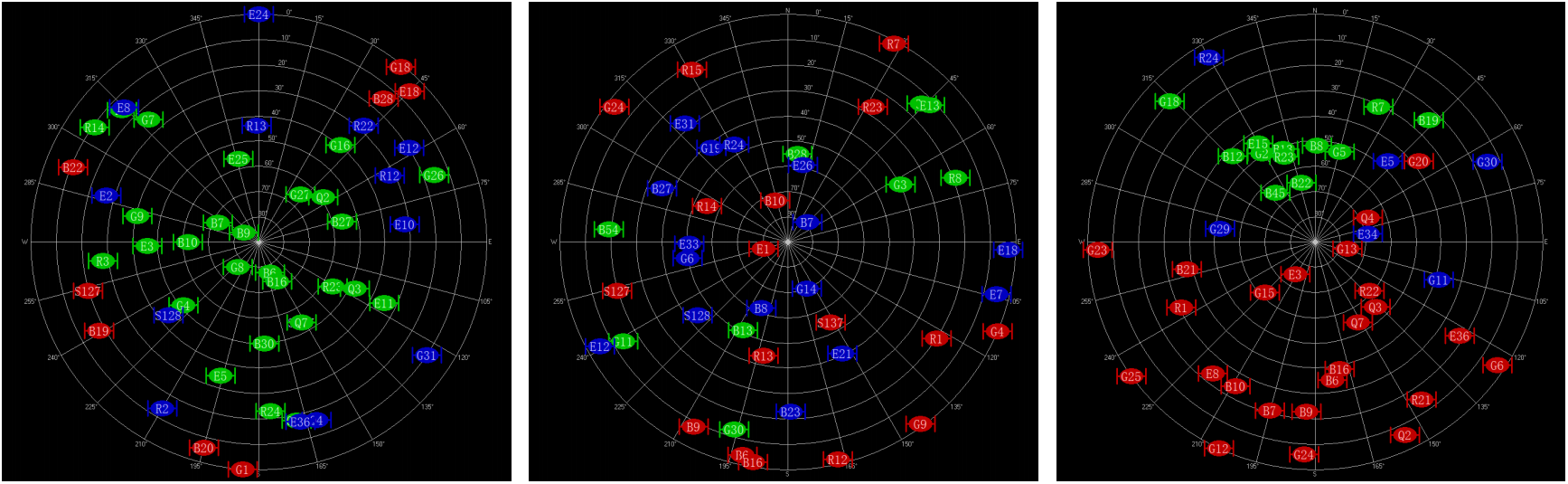}
	\caption{Satellite skyplots of open-sky (left), viaduct-down (middle) and shallow indoor (right) environments, where green represents satellites with strong $ C/N_0 $ values, blue for weak values, red for unavailable satellites.}
	\label{fig:screenshot002}
\end{figure}

Here we show the satellite skyplots (Figure \ref{fig:screenshot002}) of the GNSS signals received in three typical scenarios (open-sky, viaduct-down and shallow indoor) often encountered in CAN, with green representing satellites with strong $ C/N_0 $ values, blue for weak $ C/N_0 $ values, and red for unavailable satellites. Two key observations can be made form them: 
\begin{itemize}		
	\item[$\bullet$] Firstly, relying solely on statistical features of satellites' C/N0 and number is insufficient to effectively distinguish between each type of environment;
	\item[$\bullet$]  Secondly, there is a noticeable "semi" nature exhibited in certain environments, which can potentially be utilized as a distinguishing factor.
\end{itemize}
 The problem is how to definitely describe this "semi" nature.

\begin{figure}[h]
 	\centering
 	\includegraphics[width=.35\linewidth]{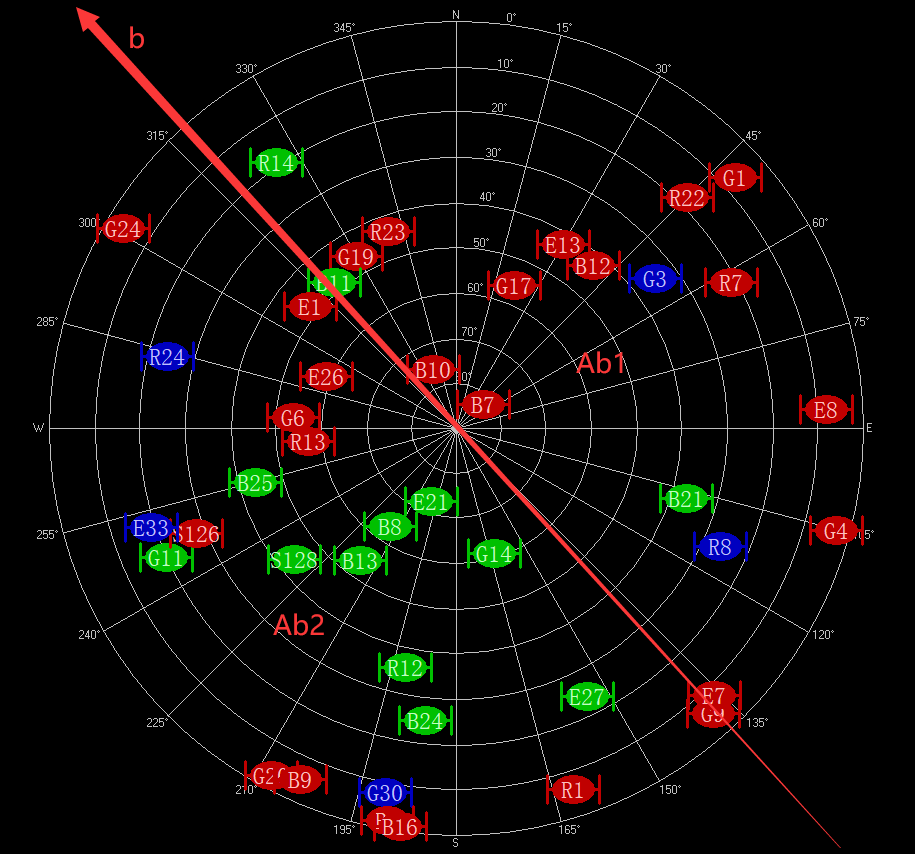}
 	\caption{Example of a satellite skyplot where $ b $ is the bisector, $ A_{b1} $ and $ A_{b2}$ are clockwise and counterclockwise ranges of $ 180^\circ  $.}
 	\label{fig:screenshot001}
 \end{figure}

 Hence, a new feature called the $ C/N_0 $-weighted azimuth distribution factor ($ r $) is designed to improve discrimination between categories.
 As shown in Figure \ref{fig:screenshot001}, we denote $b$ as the bisector. It divides the entire skyplot into two sectors, namely $ A_{b1} $ and $ A_{b2}$, respectively.
 Mathematically, $ r $ can be obtained by solving the optimization problem described in equation (\ref{key1}).
\begin{equation}\label{key1}
	\begin{split}
		&r=\max_{b \in \left[0,360^\circ\right)} \,\, r_b\\
		s.t.&\left\{\begin{array}{lc}
			r_b=\frac{CN0_{b1}}{CN0_{b2}}\\
			CN0_{bi}=\sum_{a_j\in A_{bi}}^{}c_j,i=1,2,j = 1, \ldots, N\\
			A_{b1}=\left\lbrace a \big|b\leq a<b+180^\circ~or~0^\circ\leq a<b-180^\circ \right\rbrace \\
			A_{b2}=\left\lbrace a \big|b-180^\circ\leq a<b ~or~b+180^\circ \leq a <360^\circ \right\rbrace \\
		\end{array}\right.
			\end{split}
\end{equation}
 where $a_j$ and $c_j$ denote the azimuth and $ C/N_0 $ measurement of the $j$-th available satellite,  $ A_{b1} $ and $ A_{b2}$ are clockwise and counterclockwise ranges of $ 180^\circ $ start from the bisector $ b $, respectively.
  Calculate the sum of $ C/N_0 $ of all available satellites in $ A_{b1} $ and $ A_{b2}$, respectively, and compute their ratio $ r_b $. When $ b $ varies in $ \left[0,360^\circ\right) $, the maximum value of $ r_b $ is taken as the feature of this epoch. 

 This new feature significantly improves the discrimination between different categories, especially between shallow indoor and viaduct-down. When the vehicle is in a shallow indoor environment, most of the strong GNSS signals are received from the side with windows, which will lead to a high value of $r$ (far greater than 1). On the contrary, when the vehicle is under a viaduct, the number and strength of GNSS signals from both sides are not much different, which will result in an r value close to 1. The classification model will learn this difference during the training phase and apply it during the identification phase.
 
In addition to the proposed feature $r$, the 10-dimensional feature vector used in this paper also includes the number of visible satellites, and the mean, standard deviation, maximum, minimum, skewness, kurtosis, median, and interquartile range of satellite $C/N_0$ within an epoch.
To ensure real-time performance, a lightweight GRU network is adopted for its excellent sequence data processing capabilities.   Details of the implementation of GRUs can be found in \citep{chung2014empirical}.
\section{experiments and results}
\subsection{Data Collection}
\begin{figure}[H]
		\centering
		\includegraphics[width=.85\linewidth]{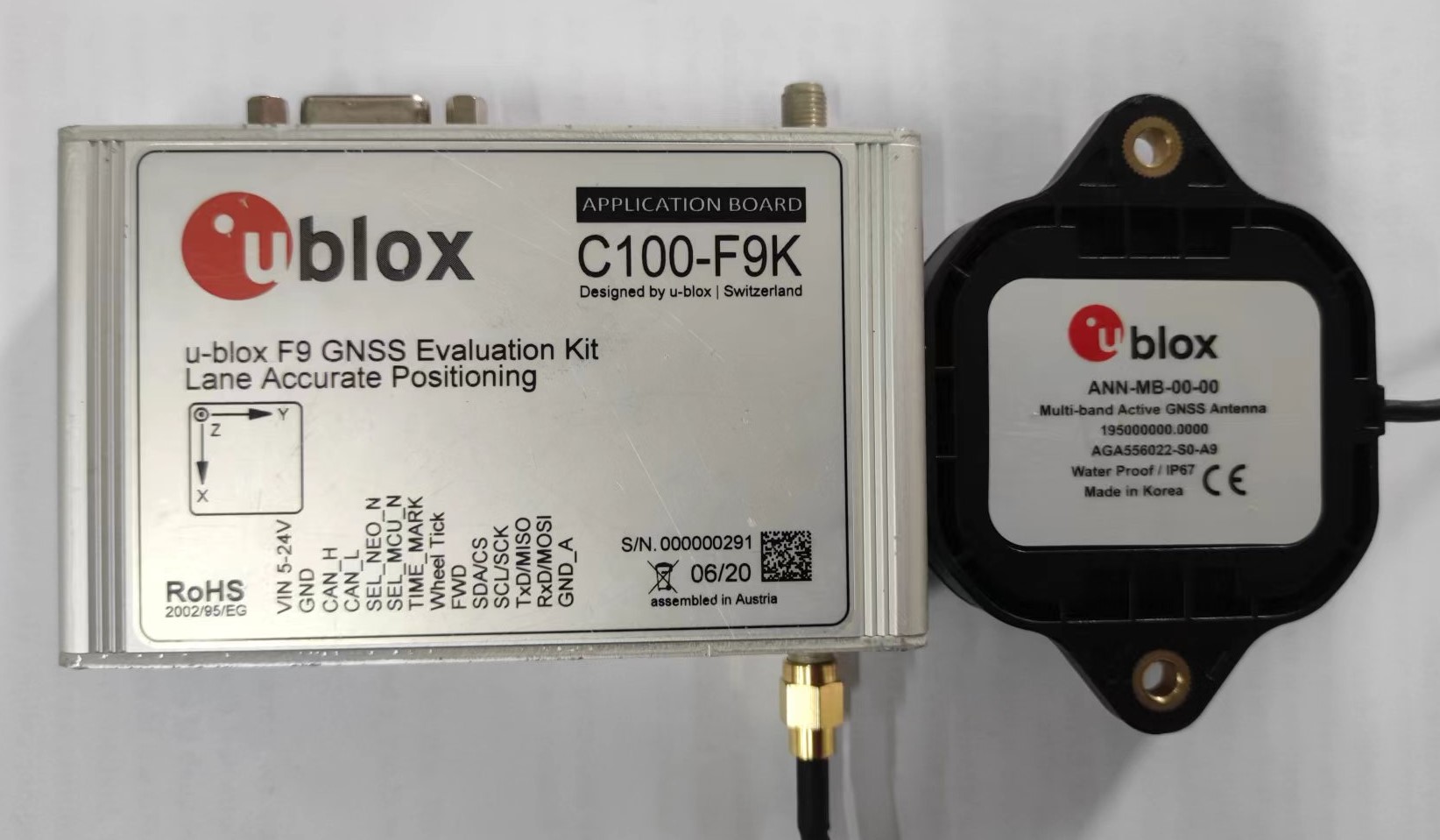}
		\caption{U-blox F9K receiver and multi-band active GNSS antenna.}
		\label{fig:20220318102324}
	\end{figure}

To validate the performance of the proposed method, extensive field experiments were conducted. 
We collected the NMEA-0183 (National Marine Electronics Association) samples using an u-blox F9K receiver (as  shown in Figure \ref{fig:20220318102324}) with a sampling frequency of 5 Hz.  Some key parameters of the used receiver are provided in Table \ref{ublox}.

The dataset consists of a total of 59,996 samples, as described in Table \ref{DATA}. Each set of data lasts approximately 4 minutes, equivalent to around 1200 samples. And to ensure the broad representativeness, they were collected by different volunteers at different locations and time for each type of environment. 
In order to facilitate further research, we have made this dataset publicly available on Github \citep{liu2020github} for researchers in the NCR community. It will serve as a benchmark for this research field.
\begin{multicols}{2}
	\setlength{\columnwidth}{1\linewidth}
	\columnbreak
	\setlength{\columnwidth}{1\linewidth}
	\begin{table}[H]
		\centering
		\caption{Parameters of the receiver U-blox F9K}
		\label{ublox}
		\begin{tabular}{@{}ll@{}}
	\toprule
	Items                & Parameters      \\ \midrule
	Constellations    & \begin{tabular}[c]{@{}l@{}}GLONASS/Galileo/ GPS/BDS/\\QZSS\end{tabular}                         \\
	Signal fequencies & \begin{tabular}[c]{@{}l@{}}L1C/A, L2C, L1OF,\\  L2OF, E1-B/C, \\ E5b, B1I, B2I\end{tabular} \\
	Max output rate      & 100 Hz          \\
	Tracking Sensitivity & -158dbm         \\
	Protocols            & NMEA, UBX, RTCM \\ \bottomrule
\end{tabular}
	\end{table}
\end{multicols}
\vspace*{-.6cm}
\begin{table}[h]
	\centering
	\caption{Number of sets and samples for each scenario.}
	\label{DATA}
	\resizebox{\columnwidth}{!}{%
		\begin{tabular}{cccccccc}
			\hline
			Categories        & open sky & tree-lined avenue & semi-outdoor & urban canyon & under-viaduct & shallow indoor & deep indoor \\ \hline
			Number of sets    & 7        & 7                 & 7            & 7            & 7             & 7              & 7           \\
			Number of samples & 8631     & 8577              & 8803         & 8507         & 8445          & 8546           & 8487        \\ \hline
		\end{tabular}%
	}
\end{table}
\subsection{Ablation Experiment}
The ablation experiment was designed to validate the performance of the proposed new feature (the $C/N_0$-weighted azimuth distribution factor $r$).  First, consider the 11-dimensional feature vector $x_t$ proposed in \citep{xia2020recurrent} :
\begin{equation}\label{key}
	x_t = \{num, sum, mean, std, max, min, range, skewness, kurtosis, median, iqr\}
\end{equation}
There are two deterministic correlations between the elements within this vector: $$mean = sum/num,range = max-min.$$
Remove $sum$ and $range$ from $x_t$, resulting in a 9-dimensional feature vector, denoted as 
\begin{equation}\label{key}
	\begin{aligned}
		y_t = \{ num,  mean, std, max, min, skewness, kurtosis, median, iqr\}.
	\end{aligned}
\end{equation}
Next, by incorporating the proposed feature $r$, the feature vector $z_t$ is obtained:
\begin{equation}\label{key}
	\begin{aligned}
		z_t = \{num,  mean, std, max, min, skewness, kurtosis, median, iqr, \underline{ratio}\}.
	\end{aligned}
\end{equation}
Then feed $y_t$ and $z_t$ into the GRU network, and evaluate their performance.
 To train a recognition model, many hyperparameters need to be set and tuned. Here we present the key parameters selected after trial and error.
 A network with two hidden layers, each containing 180 GRU neurons, was adopted. To capture the temporal correlation of scenario transitions, the length of the sliding window in the temporal domain was set to 6 samples. 
The max training epochs was set to 35, and  the batch size was set to 256, resulting a iteration number of about 7000. The learning rate was set to 5.0E-5.

\begin{figure}[h]
	\centering
	\subfloat[\scriptsize Confusion matrix (sample numbers)]{\includegraphics[width=3.20in]{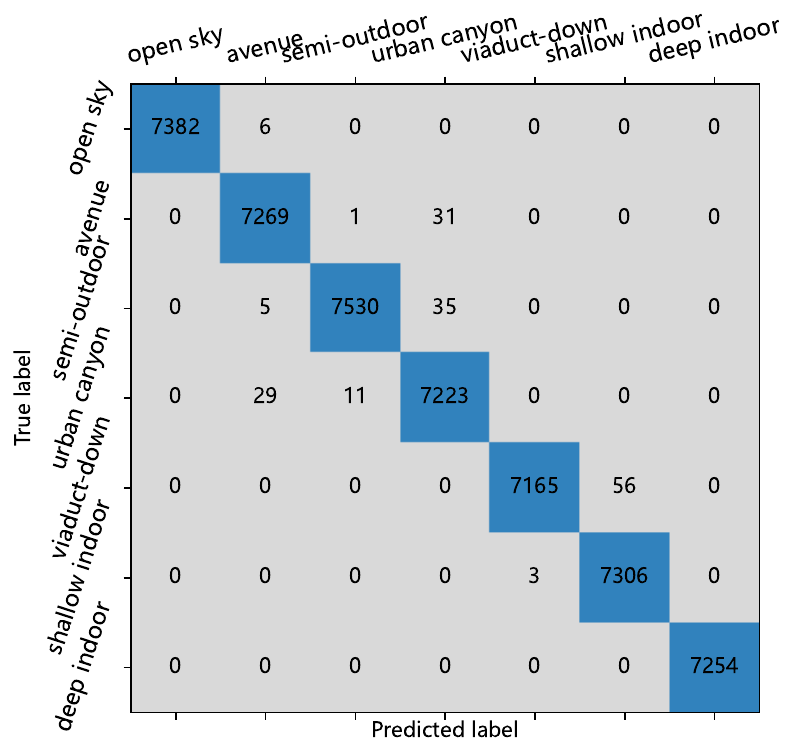}\label{subfig:a}}
	\qquad
	\subfloat[\scriptsize Confusion matrix (sample percentages)]{\includegraphics[width=3.20in]{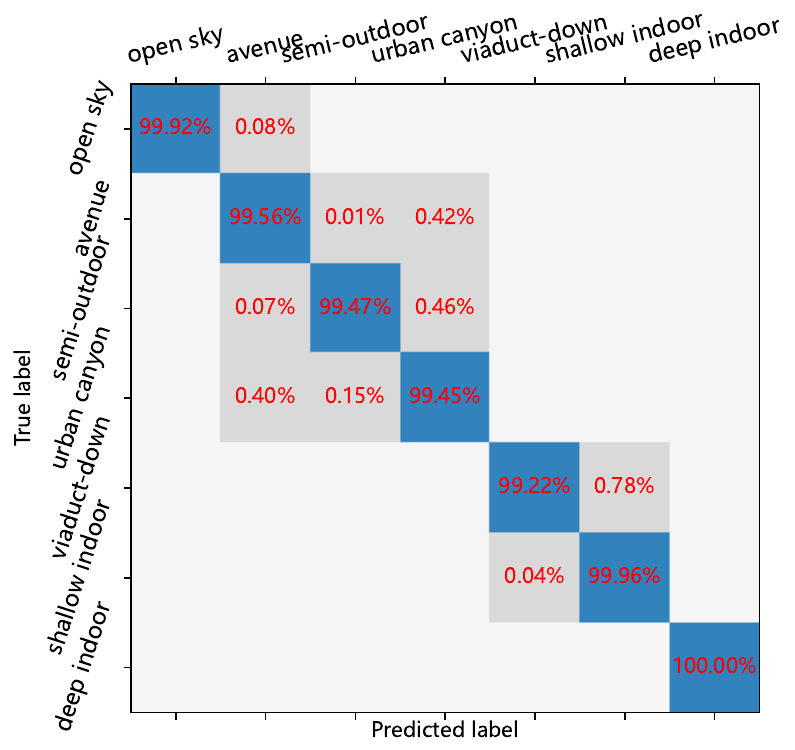}\label{subfig:b}}
	\caption{The confusion matrices of the model based on features $y_t$, overall accuracy: 99.66\%.}
	\label{confusion-y}
\end{figure}

\begin{figure}[H]
	\centering 
	\subfloat[\scriptsize Confusion matrix (sample numbers)]{\includegraphics[width=3.20in]{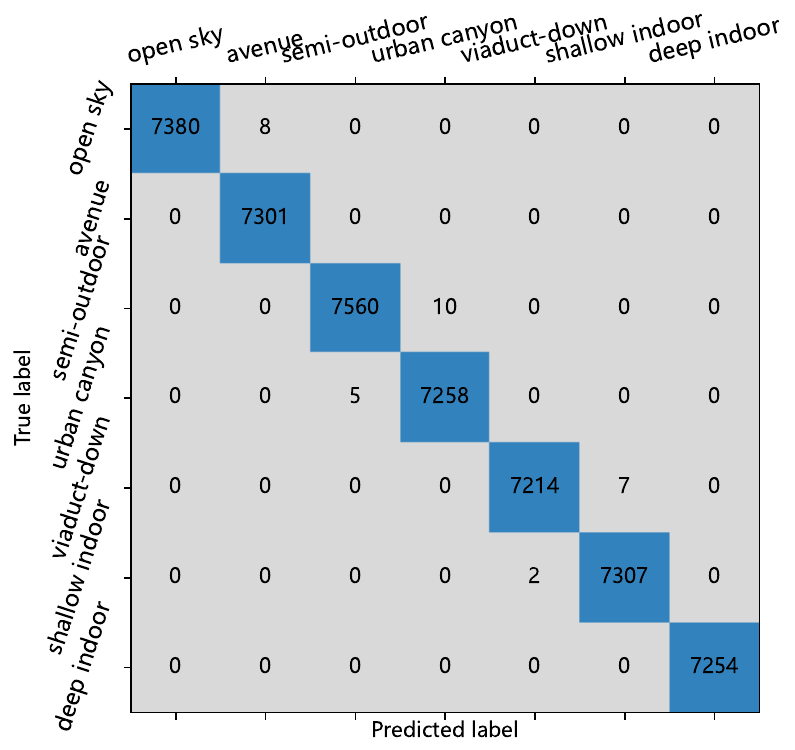}}
	\qquad 
	\subfloat[\scriptsize Confusion matrix (sample percentages)]{\includegraphics[width=3.20in]{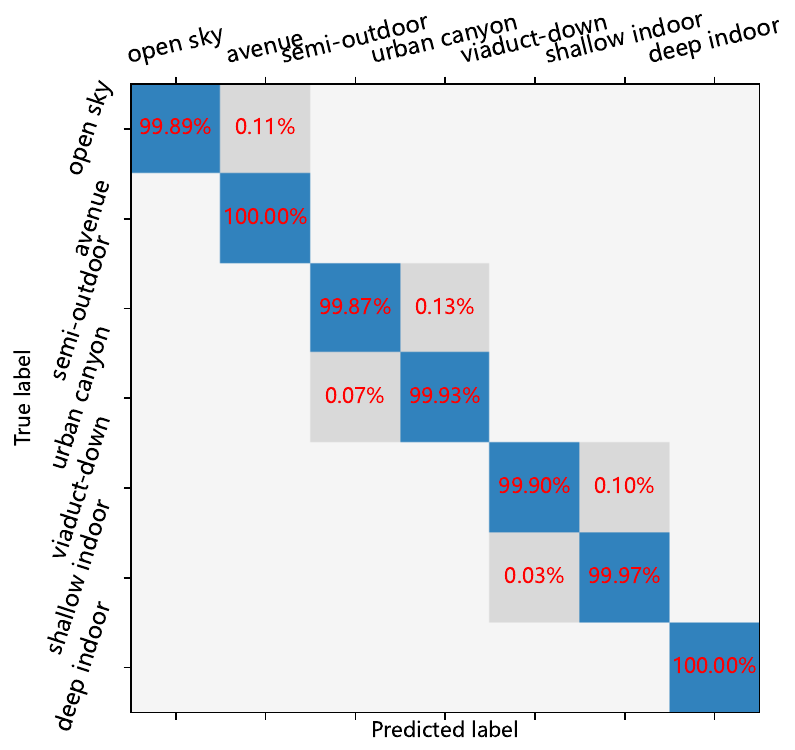}} 
	\caption{The confusion matrices of the model based on features $z_t$, overall accuracy: 99.94\% } 
	\label{confusion-z} 
\end{figure}

The classification confusion matrices of the model trained based on features $y_t$ and $z_t$ are shown in Figure \ref{confusion-y} and Figure \ref{confusion-z}, respectively. It can be observed that when trained based on $y_t$, the model tends to confuse three groups of environments. And they lead to a training accuracy of only 99.66\%. 
The most severe confusion occurs between viaduct-down and shallow indoor. When the vehicle passes between two piers under a viaduct, the semi-enclosed space obstructs the GNSS signal, reducing the number and strength of the received satellites, resulting in performance similar to entering a shallow indoor environment. However, with the new feature introduced in $z_t$, these similarities or confusions have been significantly reduced, or even eliminated altogether. As a result, we achieved an overall accuracy of 99.94\% , as shown in Fig. \ref{confusion-z}. As mentioned before, this can be attributed to the fact that distinct value of the proposed feature are obtained in the former confused environments.
\vspace{-.9cm}

\subsection{Comparison Experiments}
In the subsequent experiments, we compare the proposed GRU-based method with the SVM-based method on an isolated testset.  The SVM-based method uses temporal filtering \citep{wang2019urban} with a fixed sliding window of 6 samples and it is labeled as "SVM-TF".  They are both trained or learned using the proposed $z_t$ feature vectors. 
Their confusion matrices on the isolated test set are shown in Table \ref{tab:confusion-matrixGRU} and \ref{tab:confusion-matrixSVM}, respectively. 
The GRU-based method has a high recognition accuracy of 99.41\%, which slightly outperforms the SVM-TF-based method, 99.35\%. This can be attributed to the excellent characterization ability of the proposed $z_t$ in different environments, as well as the strong expression ability of the models.
\begin{table}[h]
	\small
	\centering
	\caption{Confusion matrix of the proposed GRU-based method on the isolated testset, overall accuracy: 99.41\%. Labels 0 to 6 represent 'open sky', ' tree-lined avenues', 'semi-outdoor', 'urban canyon', 'viaduct-down', 'shallow indoor' and 'deep indoor', respectively. The same labeling applies in the subsequent tables.}
	\begin{tabular}{|c|c|c|c|c|c|c|c|}
		\hline
		\multicolumn{1}{|c|}{\multirow{2}{*}{Actual}} & \multicolumn{7}{c|}{Predicted} \\
		\cline{2-8}
		& \multicolumn{1}{c|}{0} & \multicolumn{1}{c|}{1} & \multicolumn{1}{c|}{2} & \multicolumn{1}{c|}{3} & \multicolumn{1}{c|}{4} & \multicolumn{1}{c|}{5} & \multicolumn{1}{c|}{6} \\
		\hline
		\multicolumn{1}{|c|}{\textbf{0}} & \textbf{100.00\%} & 0.00\% & 0.00\% & 0.00\% & 0.00\% & 0.00\% & 0.00\% \\
		\multicolumn{1}{|c|}{\textbf{1}} & 0.00\% & \textbf{100.00\%} & 0.00\% & 0.00\% & 0.00\% & 0.00\% & 0.00\% \\
		\multicolumn{1}{|c|}{\textbf{2}} & 0.00\% & 0.00\% & \textbf{99.67\%} & 0.33\% & 0.00\% & 0.00\% & 0.00\% \\
		\multicolumn{1}{|c|}{\textbf{3}} & 0.00\% & 0.00\% & 0.00\% & \textbf{98.51\%} & 1.49\% & 0.00\% & 0.00\% \\
		\multicolumn{1}{|c|}{\textbf{4}} & 0.00\% & 0.00\% & 0.00\% & 0.00\% & \textbf{97.90\%} & 2.10\% & 0.00\% \\
		\multicolumn{1}{|c|}{\textbf{5}} & 0.00\% & 0.00\% & 0.25\% & 0.00\% & 0.00\% & \textbf{99.75\%} & 0.00\% \\
		\multicolumn{1}{|c|}{\textbf{6}} & 0.00\% & 0.00\% & 0.00\% & 0.00\% & 0.00\% & 0.00\% & \textbf{100.00\%} \\
		\hline
	\end{tabular}
	\label{tab:confusion-matrixGRU}
\end{table}
	\begin{table}[h]
	\small
	\centering
	\caption{Confusion matrix of the SVM-TF-based method on the isolated testset, overall accuracy: 99.35\%.}
	\begin{tabular}{|c|c|c|c|c|c|c|c|}
		\hline
		\multicolumn{1}{|c|}{\multirow{2}{*}{Actual}} & \multicolumn{7}{c|}{Predicted} \\
		\cline{2-8}
		& \multicolumn{1}{c|}{0} & \multicolumn{1}{c|}{1} & \multicolumn{1}{c|}{2} & \multicolumn{1}{c|}{3} & \multicolumn{1}{c|}{4} & \multicolumn{1}{c|}{5} & \multicolumn{1}{c|}{6} \\
		\hline
		\multicolumn{1}{|c|}{\textbf{0}} & \textbf{100.00\%} & 0.00\% & 0.00\% & 0.00\% & 0.00\% & 0.00\% & 0.00\% \\
		\multicolumn{1}{|c|}{\textbf{1}} & 0.00\% & \textbf{100.00\%} & 0.00\% & 0.00\% & 0.00\% & 0.00\% & 0.00\% \\
		\multicolumn{1}{|c|}{\textbf{2}} & 0.00\% & 0.00\% & \textbf{99.50\%} & 0.50\% & 0.00\% & 0.00\% & 0.00\% \\
		\multicolumn{1}{|c|}{\textbf{3}} & 0.00\% & 0.00\% & 0.00\% & \textbf{98.68\%} & 1.32\% & 0.00\% & 0.00\% \\
		\multicolumn{1}{|c|}{\textbf{4}} & 0.00\% & 0.00\% & 0.00\% & 0.00\% & \textbf{97.90\%} & 2.10\% & 0.00\% \\
		\multicolumn{1}{|c|}{\textbf{5}} & 0.00\% & 0.00\% & 0.67\% & 0.00\% & 0.00\% & \textbf{99.33\%} & 0.00\% \\
		\multicolumn{1}{|c|}{\textbf{6}} & 0.00\% & 0.00\% & 0.00\% & 0.00\% & 0.00\% & 0.00\% & \textbf{100.00\%} \\
		\hline
	\end{tabular}
	\label{tab:confusion-matrixSVM}
\end{table}

\begin{figure}[h]
	\centering
	\includegraphics[width=.95\linewidth]{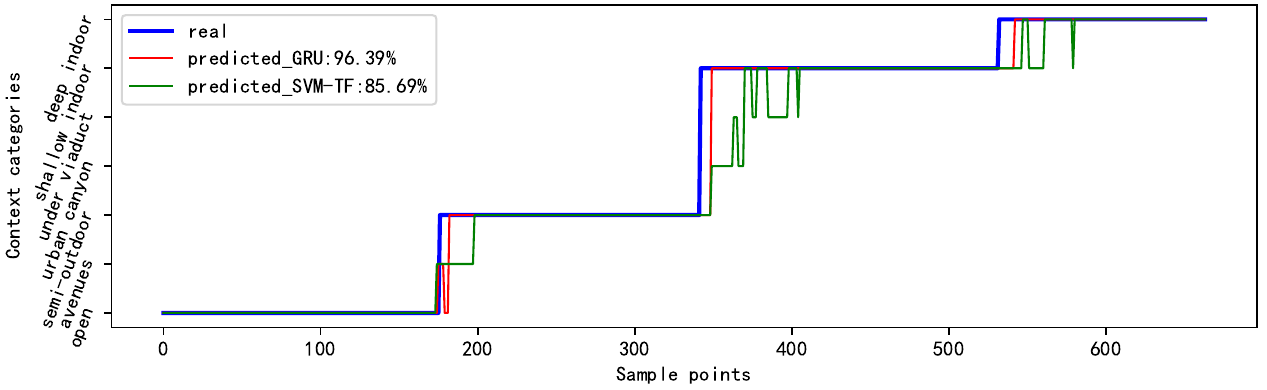}
	\caption{Transition scenario 1: open sky→semi-outdoor→shallow indoor→deep indoor}
	\label{fig:tans1}
\end{figure}

\begin{figure}[h]
	\centering
	\includegraphics[width=.95\linewidth]{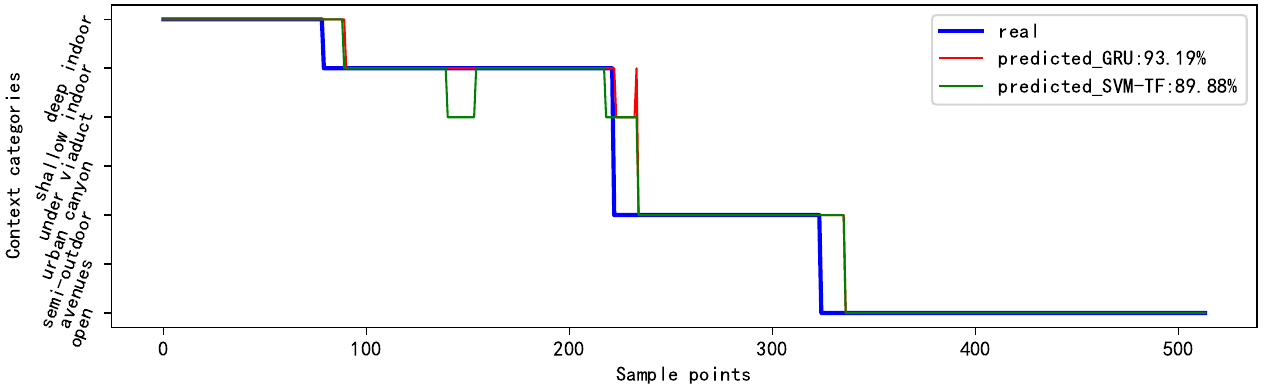}
	\caption{Transition scenario 2: deep indoor→shallow indoor→semi-outdoor→open sky}
	\label{fig:tans2}
\end{figure}

\begin{figure}[]
	\centering
	\includegraphics[width=.95\linewidth]{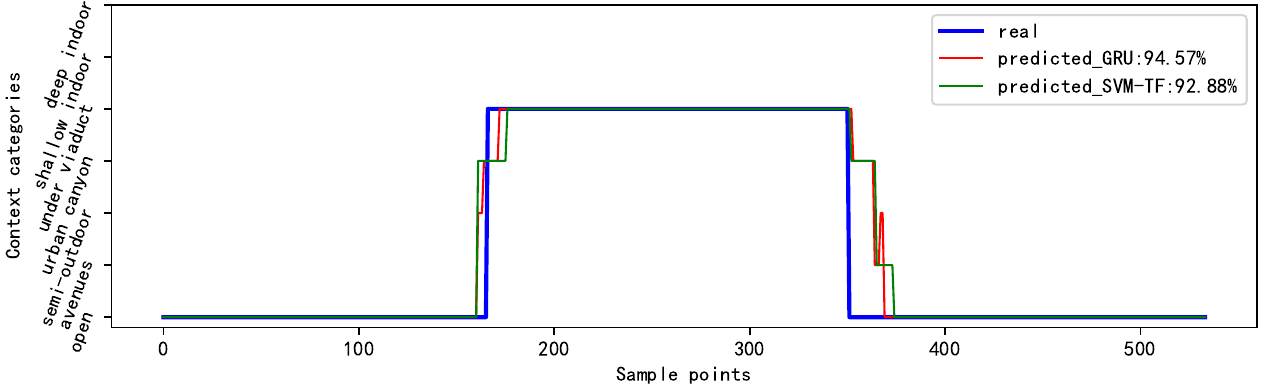}
	\caption{Transition scenario 3: open sky→viaduct-down→open sky}
	\label{fig:tans3}
\end{figure}

\begin{figure}[H]
	\centering
	\includegraphics[width=.95\linewidth]{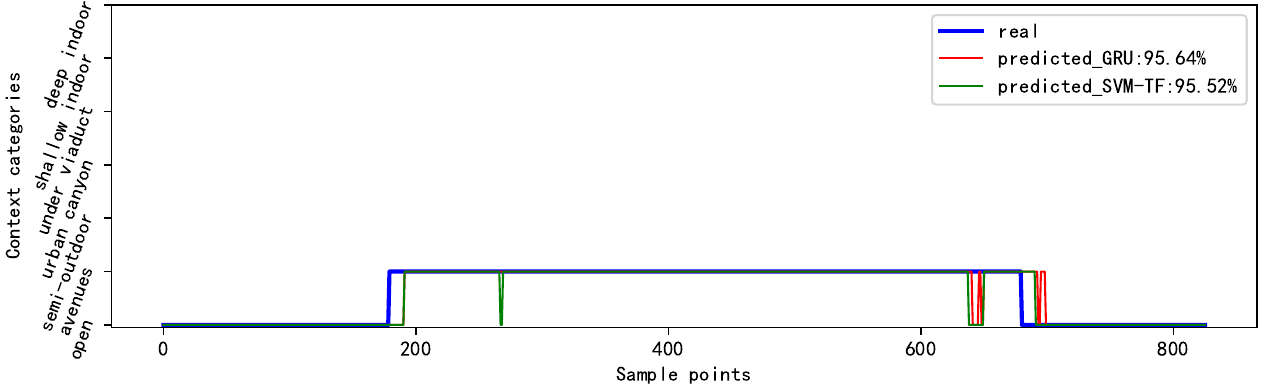}
	\caption{Transition scenario 4: open sky→tree-lined avenue→open sky}
	\label{fig:tans4}
\end{figure}

To evaluate the performance in transition scenarios, we conducted 4 sets of comparative experiments, as shown in Figure \ref{fig:tans1}-\ref{fig:tans4}.
The overall average recognition accuracy of the GRU-based method in these five transition scenarios is 94.95\%, which is significantly higher than the 90.99\% achieved by the SVM-TF-based method. This is because the former has already taken the temporal relationship of the samples into consideration during the learning stage, while the latter only applies a temporal filtering to the recognition results of single moments.
The $2^{nd}$ experiment follows the opposite process of the $1^{st}$ one.  For the GRU-based method, although both experiments are conducted in the same route, the accuracy and delay in the  $2^{nd}$ one are inferior to those in the $1^{st}$ one. This is because in the 1st experiment, the environmental changes follow a gradual decrease in skyvisibility, and the receiver can immediately detect the decrease in signal quantity and quality. Instead, in the $2^{nd}$ experiment, as the skyvisibility increases, the receiver needs some time to acquire and track the newly visible satellites. The same phenomenon can be observed in the $1^{st}$ and $2^{nd}$ halves of the $3^{rd}$ experiment, when the vehicle crossing the viaduct-down environment. Beside, there are a total of 10 transitions in these 4 experiments, with an average delays of 2.14s, demonstrating the real-time performance of the proposed method.
\section{conclusions}
In this paper, we studied the NCR approach for context-adaptive navigation. We proposed a new and fine-grained context categorization framework based on the characteristics of different environments and their corresponding integrated navigation methods, which is currently the most elaborate context categorization framework known. A new feature called the satellite azimuth distribution factor weighted by carrier-to-noise ratio $r$, was designed, which significantly improves the discrimination between different categories. To ensure real-time performance, a GRU network was adopted for its excellent sequence data processing capability. A corresponding data set was created, which will serve as a valuable resource for the NCR research community.
Experiments results show that the proposed method  is superior in terms of both recognition accuracy and computational complexity
to the state-of-the-art. Overall, this work has significant implications for the development of context-adaptive navigation systems, which have the potential to greatly enhance user experience and safety in navigation applications.

\section*{Acknowledgments}
This work was supported  in part by the National Key R\&D Program of China  under Grant No. 2020YFA0713502 and in part by Hunan Provincial Innovation Foundation For Postgraduate with Grant No. CX20210610.  We are grateful to the High Performance Computing Platform of Xiangtan University for assistance with the computations in training stage.

\printbibliography

\end{document}